\documentclass[12pt]{article}
\usepackage{axodraw}
\newcommand{\li}{\mbox{${\cal L}{\mbox i}_{2}$}} 
\newcommand{\Li}{\mbox{${\mbox L}{\mbox i}_{2}$}}
\newcommand{\gev}{\mbox{GeV}}

\newcommand{\real}{{\cal\mbox{Re\,}}}
\newcommand{\imag}{{\cal\mbox{Im\,}}}

\begin{document}
\pagestyle{empty}
\begin{flushright}
{CERN-TH/97--114}
\end{flushright}
\vspace*{5mm}
\begin{center}
{\bf NON-FACTORIZABLE CORRECTIONS TO W-PAIR PRODUCTION} \\
\vspace*{1cm} 
{\bf W.~Beenakker}$^{*)}$ \ \ ,\ \ {\bf A.P.~Chapovsky$^{\dagger)}$}\\
\vspace{0.3cm}
Instituut--Lorentz, University of Leiden, The Netherlands 
\vspace{0.6cm}\\
and 
\vspace{0.5cm}\\
{\bf F.A.~Berends} 
\vspace{0.3cm}\\
Theory Division, CERN, CH--1211 Geneva 23, Swizerland 
\vspace{0.3cm}\\ 
and 
\vspace{0.3cm}\\ 
Instituut--Lorentz, University of Leiden, The Netherlands \\
\vspace*{2cm}  
{\bf ABSTRACT} \\ \end{center}
\vspace*{5mm}
\noindent
In this paper we study the non-factorizable QED
corrections to $W$-pair-mediated (charged-current) four-fermion 
production in electron--positron collisions. A brief account of the 
obtained analytical results is given. They turn out to be different
from the ones published in the literature. For the first time
numerical results are presented, in particular the effects on the 
$W$ line-shape. These effects are of the order of a per cent. 
Applying the same methods to $ZZ$- or $ZH$-mediated four-fermion 
production, the non-factorizable ${\cal O}(\alpha)$ corrections
to the $Z$ or $H$ line-shape vanish.
 
\vspace*{1.5cm}
\begin{flushleft} CERN-TH/97--114 \\
May 1997
\end{flushleft}
\noindent 
\rule[.1in]{16.5cm}{.002in}

\noindent
$^{*)}$Research supported by a fellowship of the Royal Dutch Academy of Arts 
and Sciences.\\
$^{\dagger)}$Research supported by the Stichting FOM.
\vspace*{0.5cm}

\vfill\eject

\setcounter{page}{1}
\pagestyle{plain}

\section{Introduction}

With the start of LEP2, quantitative knowledge of the radiative corrections to
the four-fermion production process $e^+ e^- \to 4 f$
is needed \cite{review}. The full calculation of all these corrections will be
extremely involved and at present one relies on approximations \cite{review}, 
such as leading-log initial-state radiation and running couplings \cite{bhf}. 
Another approach is to exploit the fact that in particular the corrections 
associated with the production of an intermediate $W$-boson pair are important.
This (charged-current) production mechanism dominates at LEP2 energies and 
determines the LEP2 sensitivity to the mass of the $W$ boson and to the 
non-Abelian triple gauge-boson interactions. As such, one could approximate the
complete set of radiative corrections by considering only the leading terms in
an expansion around the $W$ poles. The double-pole residues thus obtained 
could be viewed as a gauge-invariant definition of corrections to ``$W$-pair 
production''. The sub-leading terms in this expansion are generically 
suppressed by powers of $\Gamma_W/M_W$, with $M_W$ and $\Gamma_W$ denoting the 
mass and width of the $W$ boson. The quality of this double-pole approximation 
degrades in the vicinity of the $W$-pair production threshold, but a few 
$\Gamma_W$ above threshold it is already quite reliable \cite{pole-scheme}. 
It is conceivable that in the near future a combination of the above-mentioned 
approximations will result in sufficiently accurate theoretical predictions 
for four-fermion production processes. 

In the double-pole approximation the complete set of first-order radiative 
corrections to the charged-current four-fermion processes can be divided into 
so-called factorizable and non-factorizable corrections 
\cite{review,pole-scheme}, i.e.~corrections that manifestly contain two 
resonant $W$ propagators and those that do not. In view of gauge-invariance
requirements, some care has to be taken with the precise definition of this 
split-up (see Sect.\,\ref{gaugeinv}).
In the factorizable corrections 
one can distinguish between corrections to $W$-pair production and $W$ decay. 
In this letter we address the question of the size of the usually neglected 
non-factorizable corrections. From the complete set of electroweak Feynman 
diagrams that contribute to the full $\cal{O}(\alpha)$ correction, we will 
therefore only consider the non-factorizable ones, both for virtual corrections
and real-photon bremsstrahlung. To be more precise, since we are only 
interested in the double-pole terms we are led to consider only  
non-factorizable QED diagrams in the soft-photon limit.
Subsequently, the photon is treated inclusively, 
without imposing any limits on the photon phase space \cite{fadin-khoze}.

This is the same approach as adopted by the authors of Ref.\,\cite{my}, who 
were the first to calculate non-factorizable $W$-pair corrections. For the 
present calculations, we have used two different methods. One is an extension 
of the 
treatment in \cite{my}, the other is a modification of the standard methods,
which involves a combination of the decomposition of multipoint scalar 
functions and the Feynman-parameter technique. The results obtained with our 
two methods are in complete mutual agreement. However, in contrast to 
\cite{my} a clear separation between virtual and real photonic
corrections has been made in both methods, which is essential to establish the 
cancellations of infrared and collinear divergences. This treatment reveals 
a significant difference between our results and those obtained by the
authors of \cite{my}.
Our final results do not contain any logarithmic terms involving the 
final-state fermion masses, whereas in the results of \cite{my} explicit 
logarithms of fermion-mass ratios occur (see discussion in Sect.\,4.1 of 
Ref.\,\cite{my}). This difference can be traced back to
the fact that although the fermion masses can formally be neglected in the
absence of collinear divergences, they have to be introduced in intermediate 
results in order to regularize those divergences before dropping out from the 
final results. 

In the next section we briefly focus on the analytical results as obtained 
with the modified standard technique. A detailed account of our study and a 
discussion of both calculational methods will be published elsewhere. 
In the last section we present numerical implications of the
non-factorizable corrections. Our calculations confirm that non-factorizable 
corrections vanish in the special case of initial--final state interference,
thereby making non-factorizable radiative corrections independent from the 
$W$ production angle, and in all cases when the integrations over both 
invariant masses of the virtual $W$ bosons are performed \cite{fadin-khoze}. 
The practical
consequence of the latter is that pure angular distributions are unaffected by
non-factorizable $\cal{O}(\alpha)$ corrections. So, the studies of non-Abelian
triple gauge-boson couplings at LEP2 \cite{anom-coupling} are not affected
by these corrections. The completely new aspect that we have addressed in our 
analysis is the effect of non-factorizable corrections on invariant-mass 
distributions ($W$ line-shapes). These distributions play a crucial role in 
extracting the $W$-boson mass from the data through direct reconstruction of 
the Breit--Wigner resonances. The non-factorizable corrections to the 
line-shapes turn out to be the same for quark and lepton final states, 
provided the integrations over the decay angles have been performed.

\section{Non-factorizable corrections: analytical results}

In this section we present the results calculated in the modified
standard technique. Details of the calculation and the 
alternative treatment, which is an extension of the method of \cite{my},
will be published elsewhere.

\subsection{Virtual corrections for purely leptonic final states}
\label{LLvirt}

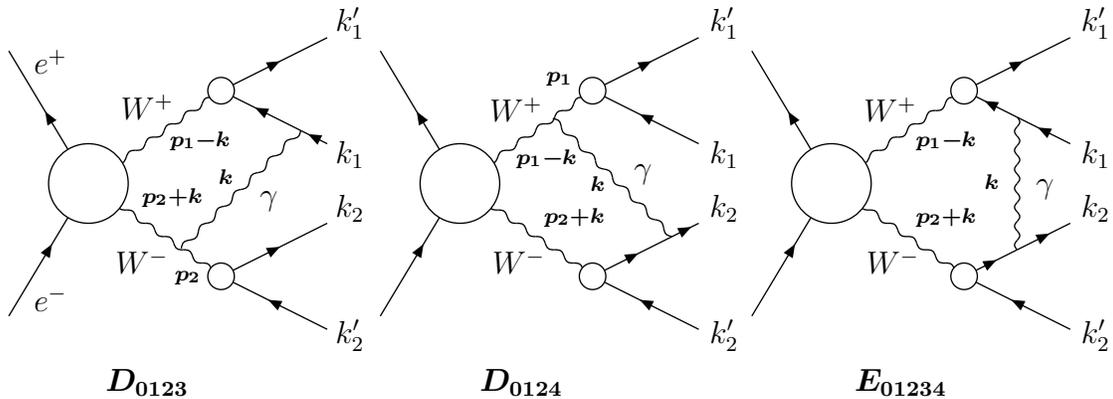
\begin{figure}
  \unitlength 0.95pt
  \begin{center}
  \begin{picture}(430,150)(0,0)
    \ArrowLine(30,75)(0,125)
    \ArrowLine(0,25)(30,75)
    \Photon(30,75)(80,110){1}{7} 
    \Photon(30,75)(80,40){1}{7}  
    \ArrowLine(120,90)(110,95) 
    \ArrowLine(110,95)(80,110) 
    \Photon(65,50.5)(110,95){1}{7}           
    \ArrowLine(80,110)(120,130)      
    \ArrowLine(120,20)(80,40)      
    \ArrowLine(80,40)(120,60)      
    \GCirc(30,75){15}{1}
    \GCirc(80,110){5}{1} 
    \GCirc(80,40){5}{1}  
    \Text(55,0)[]{\boldmath $D_{0123}$}
    \Text(10,123)[lb]{$e^+$}
    \Text(10,38)[lt]{$e^-$}
    \Text(55,105)[b]{$W^+$}
    \Text(53,53)[t]{$W^-$}
    \Text(100,76)[lt]{$\gamma$}
    \Text(130,137)[lb]{$k_1^{\prime}$}
    \Text(130,96)[lt]{$k_1$}
    \Text(130,64)[lb]{$k_2$}
    \Text(130,25)[lt]{$k_2^{\prime}$}
    \Text(89,79)[rb]{\boldmath $\scriptstyle k$}
    \Text(53,70)[lb]{\boldmath $\scriptstyle p_2+k$}
    \Text(64,100)[lt]{\boldmath $\scriptstyle p_1-k$}
    \Text(66,39)[lb]{\boldmath $\scriptstyle p_2$}
    \ArrowLine(170,75)(140,125)
    \ArrowLine(140,25)(170,75)
    \Photon(170,75)(220,110){1}{7} 
    \Photon(170,75)(220,40){1}{7}  
    \ArrowLine(260,90)(220,110)      
    \ArrowLine(220,110)(260,130)      
    \ArrowLine(260,20)(220,40)      
    \ArrowLine(220,40)(250,55)
    \ArrowLine(250,55)(260,60) 
    \Photon(205,99.5)(250,55){1}{7}     
    \GCirc(170,75){15}{1}
    \GCirc(220,110){5}{1} 
    \GCirc(220,40){5}{1} 
    \Text(204,0)[]{\boldmath $D_{0124}$} 
    \Text(202,104)[b]{$W^+$}
    \Text(203,53)[t]{$W^-$}
    \Text(249,87)[lt]{$\gamma$}
    \Text(279,137)[lb]{$k_1^{\prime}$}
    \Text(279,96)[lt]{$k_1$}
    \Text(279,64)[lb]{$k_2$}
    \Text(279,25)[lt]{$k_2^{\prime}$}
    \Text(237,82)[rt]{\boldmath $\scriptstyle k$}
    \Text(213,62)[lb]{\boldmath $\scriptstyle p_2+k$}
    \Text(202,93)[lt]{\boldmath $\scriptstyle p_1-k$}
    \Text(214,118)[lb]{\boldmath $\scriptstyle p_1$}
    \ArrowLine(310,75)(280,125)
    \ArrowLine(280,25)(310,75)
    \Photon(310,75)(360,110){1}{7} 
    \Photon(310,75)(360,40){1}{7}  
    \ArrowLine(400,90)(380,100)    
    \ArrowLine(380,100)(360,110)   
    \ArrowLine(360,110)(400,130)      
    \ArrowLine(400,20)(360,40)      
    \ArrowLine(360,40)(380,50)   
    \ArrowLine(380,50)(400,60) 
    \Photon(380,50)(380,100){1}{7}      
    \GCirc(310,75){15}{1}
    \GCirc(360,110){5}{1} 
    \GCirc(360,40){5}{1}
    \Text(355,0)[]{\boldmath $E_{01234}$}
    \Text(351,104)[b]{$W^+$}
    \Text(352,53)[t]{$W^-$}
    \Text(409,77)[l]{$\gamma$}
    \Text(427,137)[lb]{$k_1^{\prime}$}
    \Text(427,96)[lt]{$k_1$}
    \Text(427,64)[lb]{$k_2$}
    \Text(427,25)[lt]{$k_2^{\prime}$} 
    \Text(361,100)[lt]{\boldmath $\scriptstyle p_1-k$} 
    \Text(361,62)[lb]{\boldmath $\scriptstyle p_2+k$}
    \Text(394,80)[r]{\boldmath $\scriptstyle k$}
  \end{picture} 
  \caption[]{Virtual diagrams contributing to the manifestly non-factorizable 
             $W$-pair corrections. The scalar functions corresponding to these
             diagrams are denoted by $D_{0123}$, $D_{0124}$, and $E_{01234}$.}
  \label{fig:1}
  \end{center}
\end{figure}
As a first step we consider the manifestly non-factorizable corrections to the 
simplest class of charged-current four-fermion processes, involving a purely 
leptonic final state:
\begin{equation}
  e^+(q_1) e^-(q_2) \to W^+(p_1) + W^-(p_2) 
        \to \nu_{\ell}(k_1^{\prime}) \ell^+(k_1)
          + \ell^{\prime\,-}(k_2)\bar{\nu}_{\ell^{\prime}}(k_2^{\prime}).
\end{equation}
All external fermions are taken to be massless whenever possible. The relevant
contributions consist of the final--final and intermediate--final state 
photonic interactions displayed in Fig.~\ref{fig:1}. In principle also 
manifestly non-factorizable vertex corrections exist, which arise when the 
photon in Fig.~\ref{fig:1} does not originate from a $W$-boson line but from 
the $\gamma WW/ZWW$ vertex (hidden in the central blob). Those contributions 
can be shown to vanish, using power-counting arguments \cite{review}. Also the 
manifestly non-factorizable initial--final state interference effects disappear
in our approach. This happens upon adding virtual and real corrections. 

The double-pole contribution of the virtual corrections to the differential 
cross-section can be written in the form
\begin{equation}
\label{eq:mtrx_element}
  d\sigma_{\mbox{\scriptsize virt}} = 32\pi \alpha\, \real \biggl[
                         i (p_2 \cdot k_1) D_1 D_{0123}
                       + i (p_1 \cdot k_2) D_2 D_{0124} 
                       + i (k_1 \cdot k_2) D_1 D_2 E_{01234}
                       \biggr]\,d\sigma_{\mbox{\scriptsize Born}},
\end{equation}
where $D_{1,2} = p_{1,2}^2 - M_W^2 + i M_W\Gamma_W$ are the inverse 
(Breit--Wigner) $W$-boson propagators. The functions $D_{0123}$, $D_{0124}$, 
and $E_{01234}$ are the scalar integrals corresponding to the diagrams shown in
Fig.~\ref{fig:1}, with the integration measure defined as $d^4k/(2\pi)^4$.
The propagators occurring in these integrals are labelled according to:
0 = photon, 1 = $W^+$, 2 = $W^-$, 3 =  $\ell^+$, and 4 = $\ell^-$.
Note that the factorization property exhibited in Eq.\,(\ref{eq:mtrx_element})
is a direct consequence of the soft-photon approximation, which is inherent in
our approach. 

To write down the analytical results we need to introduce some 
kinematic invariants:
\begin{equation}
  m_{1,2}^2 = k_{1,2}^2, \ \ \ 
  s = (p_1 + p_2)^2, \ \ \ 
  s_{122^{\prime}} = ( k_1 + k_2 + k^{\prime}_2)^2, \ \ \ 
  s_{12}  = ( k_1 + k_2)^2,
\end{equation}
and some short-hand notations:
\begin{equation}
  y_0 = \frac{D_1}{D_2}, \ \ \ 
  \zeta = 1 - \frac{s_{122^{\prime}}}{M_W^2} - i0, \ \ \ 
  x_s = \frac{\beta-1}{\beta+1} + i0, \ \ \ 
  \beta = \sqrt{1-4M_W^2/s}\, .
\end{equation}
Here $\pm i0$ denotes an infinitesimal imaginary part, the sign of 
which is determined by causality.

The scalar four-point function $D_{0123}$ is infrared-finite, owing to the
finite decay width of the $W$ boson. In the soft-photon limit it takes the form
\begin{eqnarray}
  D_{0123} &=& \frac{i}{16\pi^2 M_W^2}\,\frac{1}{[D_2 - \zeta D_1]}
                      \Biggl\{
                          2\,\li\biggl(\frac{1}{y_0};\frac{1}{\zeta}\biggr)
                          - \li\biggl(x_s;\frac{1}{y_0}\biggr)
                          - \li\biggl(\frac{1}{x_s};\frac{1}{y_0}\biggr)
                      \Biggr. \nonumber \\[1mm]
           & &        \Biggl. {}+ \li\biggl(x_s;\zeta\biggr) 
                          + \li\biggl(\frac{1}{x_s};\zeta\biggr)
                          + \biggl[ \ln\biggl(\frac{M_W^2}{m_1^2}\biggr)
                                    \!+\! 2\ln(\zeta) \biggr]
                            \biggl[ \ln(y_0) \!+\! \ln(\zeta)\biggr] 
                      \Biggr\}.
\end{eqnarray}
The function $\li(x;y)$ is the continued dilogarithm
\begin{equation}
  \li(x;y) = \Li(1-xy) + \ln(1-xy)\,[\ln(xy)-\ln(x)-\ln(y)],
\end{equation}
with $\Li(x)$ the usual dilogarithm and $x,y$ lying on the first Riemann sheet.
The answer for the second four-point function, $D_{0124}$, can be written in a 
similar way.

The five-point scalar function, $E_{01234}$, can be evaluated by means of a
decomposition into a sum of four-point functions \cite{neerven}. In the 
double-pole approximation this decomposition reads
\begin{equation}
\label{eq:decomposition}
  w^2\,E_{01234} = 2 \Delta_{4}\,D_{1234}   
                 + (w \cdot v_1)\,D_{0234} + (w \cdot v_2)\,D_{0134} 
                 + (w \cdot v_3)\,D_{0124} + (w \cdot v_4)\,D_{0123},
\end{equation}
with 
\begin{eqnarray} 
          v_{1 \mu} &=& -\, \varepsilon_{\mu \alpha \beta \gamma}\, 
                          p_2^{\alpha} k_1^{\beta} k_2^{\gamma},
          \ \ \ 
          v_{2 \mu}  =  +\, \varepsilon_{\alpha \mu \beta \gamma}\, 
                          p_1^{\alpha} k_1^{\beta} k_2^{\gamma},
          \ \ \ 
          v_{3 \mu}  =  -\, \varepsilon_{\alpha \beta \mu \gamma}\, 
                          p_1^{\alpha} p_2^{\beta} k_2^{\gamma},
          \nonumber \\
          v_{4 \mu} &=& +\, \varepsilon_{\alpha \beta \gamma \mu}\, 
                          p_1^{\alpha} p_2^{\beta} k_1^{\gamma},
          \ \ \ 
          w^{\mu}    =    D_1 v_1^{\mu} + D_2 v_2^{\mu},
          \ \ \ \ \ \ \;
          \Delta_4 = [\, \varepsilon_{\alpha \beta \gamma \delta}\, 
                       p_1^{\alpha}p_2^{\beta} k_1^{\gamma} k_2^{\delta}\, ]^2,
\end{eqnarray}
using the convention $\varepsilon^{0123} = -\varepsilon_{0123} = 1$.
The labelling of the scalar functions ($D_{ijkl}$) is defined below
Eq.\,(\ref{eq:mtrx_element}). Note that the scalar four-point function 
$D_{1234}$ is purely a consequence of the decomposition 
(\ref{eq:decomposition}). It does not involve the exchange of a photon and is 
therefore not affected by the soft-photon approximation. 
Since we are only interested in the double-pole residue, it should be 
calculated for on-shell $W$ bosons. For the analytical 
expression, which is too involved to be presented here, we refer
to the literature \cite{gj}. The other new scalar four-point functions, 
$D_{0134}$ and $D_{0234}$, are infrared-divergent and should be calculated in 
the soft-photon approximation. Using a regulator mass $\lambda$ for the
photon we can write
\begin{eqnarray}
  D_{0234} &=& - \frac{i}{16 \pi^2 s_{12}}\,\frac{1}{D_2}
               \Biggl[ \Li\biggl(1 + \frac{\zeta M_W^2}{s_{12}}\biggr)
                       - 2 \ln \biggl( \frac{M_W \lambda}{-D_2} \biggr)
                           \ln \biggl( \frac{m_1 m_2}{-s_{12}-i0} \biggr)
               \Biggr. \nonumber \\[1mm]
           & & \Biggl. \hphantom{- \frac{i}{16 \pi^2 s_{12}}\,\frac{1}{D_2}a}
                       + \frac{\pi^2}{3}
                       + \ln^2\biggl( \frac{M_W}{m_2} \biggr)
                       + \ln^2\biggl( \frac{m_1}{\zeta M_W} \biggr)
               \Biggr],
\end{eqnarray}
with a similar expression for $D_{0134}$.

\subsection{Real-photon radiation for purely leptonic final states}
\label{sec:LL/real}

Only interferences of the real-photon diagrams can give contributions to the
manifestly non-factorizable corrections. The relevant interferences can be 
read off from
Fig.~\ref{fig:1} by taking the exchanged photon to be on-shell. The infrared 
divergences contained in the virtual corrections will cancel against those 
present in the corresponding bremsstrahlung interferences.

It should be noted that it is more complicated to obtain the five-point 
radiative interference correction. This is because the 
decomposition that we used in the case of the virtual five-point function 
cannot be carried over to the real-photon case. However, it is still possible 
to derive another decomposition using similar, but less straightforward 
arguments. Denoting the radiative analogues of the virtual scalar functions by
a superscript `R', we find
\begin{eqnarray}
\label{eq:realdecomp}
  w^{\prime 2}\,E^{\mbox{\scriptsize R}}_{01234} &=& 
       (w^{\prime} \cdot v_1^{\prime})\,D^{\mbox{\scriptsize R}}_{0234} 
     + (w^{\prime} \cdot v_2^{\prime})\,D^{\mbox{\scriptsize R}}_{0134} 
     + (w^{\prime} \cdot v_3^{\prime})\,D^{\mbox{\scriptsize R}}_{0124} 
  \nonumber \\[2mm]
                                                 & & {}
     + (w^{\prime} \cdot v_4^{\prime})\,D^{\mbox{\scriptsize R}}_{0123}
     + 2i \Delta_{4}\,D^{\mbox{\scriptsize R}}_{1234}.
\end{eqnarray}
The four-vectors $w^{\prime}$ and $v_{i}^{\prime}$ are defined as before, but 
for real-photon emission. This is equivalent to the following substitutions: 
$p_1 \to -p_1$, $k_1 \to -k_1$ and $D_2 \to D_2^*$. The radiation function
$D_{1234}^{\mbox{\scriptsize R}}$ is an artefact of the decomposition 
(\ref{eq:realdecomp}) and 
does not involve the exchange of a photon. It can be obtained from $D_{1234}$
by the substitutions $p_1 \to -p_1$ and $k_1 \to -k_1$, resulting in the 
relation $\imag D_{1234}^{\mbox{\scriptsize R}} = \imag D_{1234}$.

As will be explained in detail elsewhere, the radiative interferences can in 
fact be obtained from the virtual corrections by only considering the 
contribution from the photon pole to the complex $k^{0}$ integration
and by making certain substitutions. The photon-pole part $D^{\gamma}_{ijkl}$ 
of the scalar four-point function $D_{ijkl}$ is obtained by subtracting the 
particle-pole part $D^{\mbox{\scriptsize part}}_{ijkl}$ from $D_{ijkl}$:
\begin{equation}
  D^{\gamma}_{ijkl} = D_{ijkl} - D^{\mbox{\scriptsize part}}_{ijkl}.
\end{equation}
The particle-pole parts are found to be
\begin{eqnarray}             
  D_{0123}^{\mbox{\scriptsize part}} &=& 
                      \frac{1}{8 \pi M_W^2}\,\frac{1}{D_2 - \zeta D_1}
                      \Biggl[ \ln\bigl( 1 - y_0 x_s \bigr)
                            - \ln\bigl( 1 - x_s/\zeta \bigr)
                      \Biggr], \\[1mm]
  D_{0234}^{\mbox{\scriptsize part}} &=& 
                      \frac{1}{8\pi s_{12}}\,\frac{1}{D_2}
                      \Biggl[ \ln\biggl( \frac{D_2}{i M_W^2} \biggr)
                            - \ln(-\zeta)
                            - \ln\biggl( \frac{\lambda}{m_1} \biggr)
                      \Biggr],
\end{eqnarray}
with similar expressions for $D_{0124}^{\mbox{\scriptsize part}}$ and 
$D_{0134}^{\mbox{\scriptsize part}}$, respectively. The radiative 
interferences can be obtained from Eq.\,(\ref{eq:mtrx_element}) by adding a 
minus sign, by inserting the decomposition given in 
Eq.\,(\ref{eq:decomposition}), and by substituting 
\begin{eqnarray*}
  \mbox{-- in the $D_{0123},D_{0134}$ terms:} && \!\!\!\! 
       D_{0123},D_{0134} \to D^{\gamma}_{0123},D^{\gamma}_{0134}
       \mbox{ \ followed by \ } D_1 \to -D_1^*, \\
  \mbox{-- in the $D_{0124},D_{0234}$ terms:} &&  \!\!\!\! 
       D_{0124},D_{0234} \to D^{\gamma}_{0124},D^{\gamma}_{0234}
       \mbox{ \ followed by \ } D_2 \to -D_2^*, \\
  \mbox{-- in the $D_{1234}$ terms:}\hphantom{,D_{0234}} &&  \!\!\!\! 
       D_{1234} \to D^{\mbox{\scriptsize R}}_{1234}
       \hphantom{,D^{\gamma}_{0124},D^{\gamma}_{0234}} 
       \mbox{ \ followed by \ } D_2 \to -D_2^*. 
\end{eqnarray*}

\subsection{Gauge-invariant definition of non-factorizable corrections}
\label{gaugeinv}

The set of manifestly non-factorizable QED diagrams displayed in 
Fig.~\ref{fig:1} is not gauge invariant. In order to achieve a gauge-invariant
definition of the non-factorizable corrections, all (soft) photonic 
interactions between the positively ($e^+,W^+,\ell^+$) and negatively
($e^-,W^-,\ell^{\prime\, -}$) charged particles should be taken into account.
Looking at Fig.~\ref{fig:1}, this is equivalent to the set of all 
up--down QED interferences. In the soft-photon, double-pole approximation only
the ``Coulomb'' interaction between the off-shell $W$ bosons survives as an 
extra contribution to the differential cross-section:
\begin{equation}
\label{Coulvirt}
  d\sigma^{\mbox{\scriptsize C}}_{\mbox{\scriptsize virt}}(p_1|p_2) =
  32\pi \alpha\, \real \biggl[ i (p_1 \cdot p_2) C_{012} \biggr]\,
  d\sigma_{\mbox{\scriptsize Born}}.
\end{equation}
The scalar three-point function $C_{012}$ is defined according to the notation
of Sect.\,\ref{LLvirt}. In our approximation it is artificially 
ultraviolet-divergent. Introducing an upper bound $\Lambda$ for the energy of 
the photon, this scalar function reads
\begin{eqnarray}
\label{C012}
  C_{012} &=& \frac{i}{16\pi^2 s \beta} \Biggl\{ 
                \li\biggl( y_0;\frac{1}{x_s} \biggr)
              + \li\biggl( \frac{1}{y_0};\frac{1}{x_s} \biggr) 
              - 2\,\Li\biggl( 1-\frac{1}{x_s} \biggr) 
              + \frac{1}{2}\,\ln^2(y_0) \Biggr. \nonumber \\[1mm]
          & &                           \Biggl. {}
              + \ln(x_s) \Biggl[ \ln\biggl( \frac{-iD_1}{2M_W \Lambda} \biggr)
                + \ln\biggl( \frac{-iD_2}{2M_W \Lambda} \biggr) \Biggr]
              - 2i\pi\,\ln\biggl( \frac{1+x_s}{2} \biggr) 
                                        \Biggr\}.
\end{eqnarray}
The corresponding radiative interference can again be related to the virtual
correction (\ref{Coulvirt}) by adding a minus sign and by substituting 
$C_{012} \to C_{012}^{\gamma}$ followed by $D_1 \to - D_1^*$. The photon-pole
part $C_{012}^{\gamma} = C_{012} - C_{012}^{\mbox{\scriptsize part}}$ can be
derived from Eq.\,(\ref{C012}) and 
\begin{equation}
  C_{012}^{\mbox{\scriptsize part}} = \frac{1}{8\pi s \beta} \Biggl\{
              \ln(1-x_s) + \ln(1+x_s) - \ln(1-y_0 x_s)
              - \ln\biggl( \frac{-iD_2}{M_W \Lambda} \biggr) \Biggr\}.
\end{equation}
If the virtual and real corrections are added, the dependence on the cut-off
parameter $\Lambda$ vanishes. When we mention {\em non-factorizable}
corrections in the following, we implicitly refer to the gauge-invariant sum
of the manifestly non-factorizable corrections and the above-mentioned 
``Coulomb'' contribution.

\subsection{Semi-leptonic and purely hadronic final states}

For the purely hadronic final states there are many more diagrams, as the
photon can interact with all four final-state fermions. In order to make
efficient use of the results presented in the previous subsections, we first 
introduce some short-hand notations based on the results for the purely 
leptonic~($LL$) final states. These short-hand notations involve the summation 
of virtual and real corrections to the differential cross-section. 
For instance, the virtual corrections originating from the first diagram of 
Fig.~\ref{fig:1} can be combined with the corresponding real-photon correction 
into the contribution $d\sigma_{LL}^{(4)}(k_1;k_1^{\prime}|p_2)$.
In a similar way virtual and real five-point corrections can be combined into
$d\sigma_{LL}^{(5)}(k_1;k_1^{\prime}|k_2;k_2^{\prime})$. The gauge-restoring
``Coulomb'' contribution will be indicated by 
$d\sigma^{\mbox{\scriptsize C}}(p_1|p_2)$. In terms of this notation the 
non-factorizable differential cross-section for purely leptonic final states 
becomes
\begin{equation}
\label{eq:LL}
  d\sigma_{LL}(k_1;k_1^{\prime}|k_2;k_2^{\prime}) = 
                    d\sigma_{LL}^{(4)}(k_1;k_1^{\prime}|p_2)
                  + d\sigma_{LL}^{(4)}(k_2;k_2^{\prime}|p_1)
                  + d\sigma_{LL}^{(5)}(k_1;k_1^{\prime}|k_2;k_2^{\prime})
                  + d\sigma^{\mbox{\scriptsize C}}(p_1|p_2).
\end{equation}
Analogously the non-factorizable differential cross-section for a purely 
hadronic final state~($HH$) can be written in the following way 
\begin{eqnarray}
\label{eq:HH}
  \lefteqn{\hspace*{-6mm}d\sigma_{HH}(k_1;k_1^{\prime}|k_2;k_2^{\prime}) = 
        3\times 3 \biggl[
        \frac{1}{3}\,d\sigma_{LL}^{(4)}(k_1;k_1^{\prime}|p_2)
      + \frac{2}{3}\,d\sigma_{LL}^{(4)}(k_1^{\prime};k_1|p_2)
      + \frac{1}{3}\,d\sigma_{LL}^{(4)}(k_2;k_2^{\prime}|p_1)}
                  \biggr. \nonumber \\[1mm]
              & & \biggl.
    {}+ \frac{2}{3\,}\,d\sigma_{LL}^{(4)}(k_2^{\prime};k_2|p_1)
      + \frac{1}{3}\cdot \frac{1}{3}\,
        d\sigma_{LL}^{(5)}(k_1;k_1^{\prime}|k_2;k_2^{\prime})
      + \frac{2}{3}\cdot \frac{1}{3}\,  
        d\sigma_{LL}^{(5)}(k_1^{\prime};k_1|k_2;k_2^{\prime})
                  \biggr. \nonumber \\[1mm]
              & & \biggl.
    {}+ \frac{1}{3}\cdot \frac{2}{3}\,   
        d\sigma_{LL}^{(5)}(k_1;k_1^{\prime}|k_2^{\prime};k_2)
      + \frac{2}{3}\cdot \frac{2}{3}\,   
        d\sigma_{LL}^{(5)}(k_1^{\prime};k_1|k_2^{\prime};k_2)
      + d\sigma^{\mbox{\scriptsize C}}(p_1|p_2)
  \biggr].
\end{eqnarray}
In order to keep the notation as uniform as possible, the momenta of the 
final-state quarks are defined along the lines of the purely leptonic case 
with $k_i$ ($k^{\prime}_i$) corresponding to down (up) type quarks. If one 
would like to take into account quark-mixing effects, it suffices to add the 
appropriate squared quark-mixing matrix elements ($|V_{ij}|^2$) to the overall 
factor. Note that top quarks do not contribute to the double-pole residues, 
since the on-shell decay $W \to t b$ is not allowed. Therefore the 
approximation of massless final-state fermions is still justified. 

For a semi-leptonic final state (say $HL$), when the $W^+$ decays 
hadronically and the $W^-$ leptonically, one can write
\begin{eqnarray}
\label{eq:HL}
    \lefteqn{\hspace*{-6mm} d\sigma_{HL}(k_1;k_1^{\prime}|k_2;k_2^{\prime}) =
             3 \biggl[ 
               \frac{1}{3}\,d\sigma_{LL}^{(4)}(k_1;k_1^{\prime}|p_2)
             + \frac{2}{3}\,d\sigma_{LL}^{(4)}(k_1^{\prime};k_1|p_2)
             +              d\sigma_{LL}^{(4)}(k_2;k_2^{\prime}|p_1)}
               \biggr. \nonumber \\
                                                  & &   \biggl. {}
      + \frac{1}{3}\,d\sigma_{LL}^{(5)}(k_1;k_1^{\prime}|k_2;k_2^{\prime})
      + \frac{2}{3}\,d\sigma_{LL}^{(5)}(k_1^{\prime};k_1|k_2;k_2^{\prime})
      + d\sigma^{\mbox{\scriptsize C}}(p_1|p_2)
                                                        \biggr].
\end{eqnarray}

Upon integration over the decay angles, the functions $d\sigma_{LL}^{(5)}$
and $d\sigma_{LL}^{(4)}$ become symmetric under 
$k_i \leftrightarrow k_i^{\prime}$. As a result, the expressions 
(\ref{eq:HH}) and (\ref{eq:HL}) take on the form of (\ref{eq:LL}) multiplied
by the colour factors 9 and 3, respectively. These are precisely the colour
factors that also arise in the Born cross-section. Therefore, after
integration over the decay angles, the relative non-factorizable correction is
the same for all final states. This universality property holds for all 
situations that exhibit the $k_i \leftrightarrow k_i^{\prime}$ symmetry.

\section{Numerical results}

\begin{figure}[t]
  \unitlength 1cm
  \begin{center}
  \begin{picture}(13.4,7)
  \put(1.2,5.8){\makebox[0pt][c]{\boldmath $\delta_{nf}$}}
  \put(11.5,-0.3){\makebox[0pt][c]{\boldmath $\cos\theta_1$}}
  \put(0,-5.5){\includegraphics{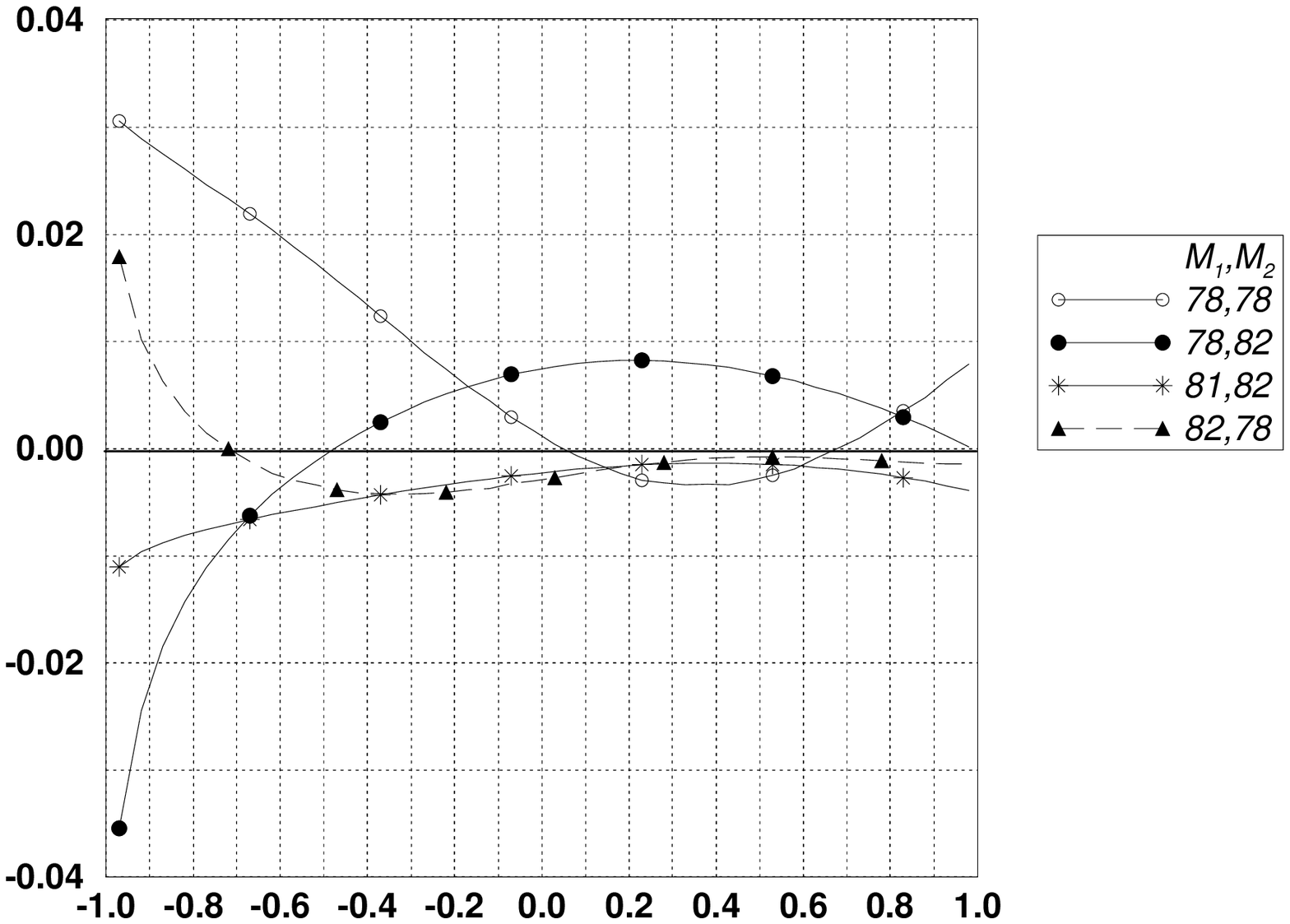}}
  \end{picture}
  \end{center}
  \caption[]{The relative non-factorizable correction 
             $\delta_{nf}(M_1,M_2,\cos\theta_1)$ to the decay angular 
             distribution $d\sigma/[dM_1 dM_2 d\cos\theta_1]$
             for fixed values of the invariant masses $M_{1,2}$ [in GeV]. 
             Centre-of-mass energy: $\sqrt{s}=184~\gev$.}
\label{fig:2}
\end{figure}

\begin{figure}[htb]
  \unitlength 1cm
  \begin{center}
  \begin{picture}(13.4,8)
  \put(7.1,7.1){\makebox[0pt][c]{\boldmath $M_{W}$}}
  \put(7.02,6.75){\makebox[0pt][c]{\boldmath $\downarrow$}}
  \put(1.8,6.1){\makebox[0pt][c]{\boldmath $\delta_{nf}$}}
  \put(12.3,-0.3){\makebox[0pt][c]{\boldmath $M$ \bf [GeV]}}
  \put(1,-5.5){\includegraphics{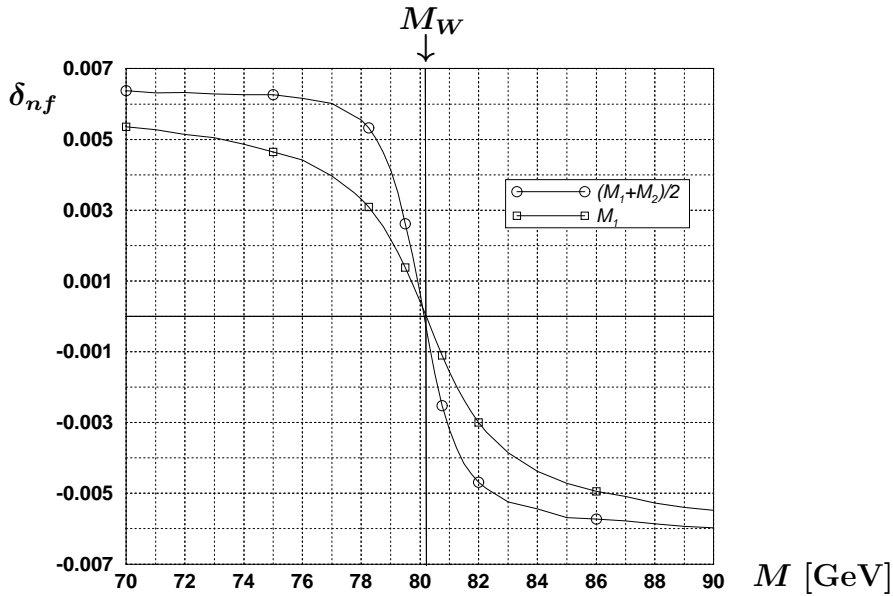}}
  \end{picture}
  \end{center}
  \caption[]{The relative non-factorizable correction $\delta_{nf}(M)$ to the 
             single invariant-mass distributions $d\sigma/dM$. Shown are the
             corrections to the distributions with respect to $M_1$ and 
             $M_{av}$. Centre-of-mass energy: $\sqrt{s}=184~\gev$.}
\label{fig:3}
\end{figure}

In this section some numerical results will be presented.
The quantity of interest is the relative non-factorizable 
correction $\delta_{nf}$, defined as
\begin{equation}
  \frac{d\sigma}{d\xi} = \frac{d\sigma_{\mbox{\scriptsize Born}}}{d\xi}\,
                         \bigl[1 + \delta_{nf}(\xi)\bigr],
\end{equation}
where $\xi$ represents some set of variables. Here we consider consecutively
the distributions $d\sigma/[dM_1 dM_2 d\cos\theta_1]$, 
$d\sigma/[dM_1 dM_2]$, $d\sigma/dM_1$ and $d\sigma/dM_{av}$, with 
$M_i= \sqrt{p_i^2}\,$, $M_{av}=\frac{1}{2}(M_1+M_2)$ and $\theta_1$ is the 
decay angle between $\vec{k}_1$ and $\vec{p}_1$ in the \underline{lab} system.
The results are shown in Fig.~\ref{fig:2} for the angular distribution, and in
Table~\ref{tab:1} and Fig.~\ref{fig:3} for the invariant-mass distributions.
The pure invariant-mass distributions play an important
role in the extraction of the $W$-boson mass from the data through direct 
reconstruction of the Breit--Wigner resonances. In this context especially the
position of the maximum of these Breit--Wigner curves is of importance.%
\begin{table}[b]
\[
  \begin{array}{||c||c|c|c|c|c|c|c||}    \hline \hline
    \raisebox{-3mm}{$\Delta_1$} 
    & \multicolumn{7}{|c||}{\raisebox{-1mm}{$\Delta_2$}}\\ 
    \cline{2-8}
         & -1    & -1/2  & -1/4  &  0    & 1/4   & 1/2   & 1     \\ 
    \hline  \hline
    -1   &+0.84 &+0.65 &+0.51 &+0.37 &+0.21 &+0.06 &-0.17 \\ 
    -1/2 &+0.65 &+0.52 &+0.39 &+0.24 &+0.07 &-0.07 &-0.26 \\
    -1/4 &+0.51 &+0.39 &+0.28 &+0.13 &-0.02 &-0.15 &-0.31 \\ 
     0   &+0.37 &+0.24 &+0.13 &+0.00 &-0.13 &-0.24 &-0.37 \\ 
     1/4 &+0.21 &+0.07 &-0.02 &-0.13 &-0.24 &-0.32 &-0.43 \\ 
     1/2 &+0.06 &-0.07 &-0.15 &-0.24 &-0.32 &-0.39 &-0.47 \\ 
     1   &-0.17 &-0.26 &-0.31 &-0.37 &-0.43 &-0.47 &-0.53 \\ 
    \hline \hline
  \end{array}
\]
\caption[]{The relative non-factorizable correction $\delta_{nf}(M_1,M_2)$ 
           [in \%] to the double invariant-mass distribution 
           $d\sigma/[dM_1 dM_2]$ for some particular values of $M_{1,2}$. 
           The invariant masses $M_{1,2}$ are specified in terms of their 
           distance from $M_W$ in units of $\Gamma_W$, 
           i.e.~$\Delta_{1,2} = [M_{1,2}-M_W]/\Gamma_W$. 
           Centre-of-mass energy: $\sqrt{s}=184~\gev$.}
\label{tab:1}
\end{table}%

All results in this section are presented for the following set of input
parameters: 
$$
  M_{W}=80.22~\gev,  \ \ \ \Gamma_{W}=2.08~\gev, \ \ \     
  M_{Z}=91.187~\gev, \ \ \ \Gamma_{Z}=2.49~\gev,
$$
$$
  \alpha=1/137.0359895, \ \ \ \sin^{2}\theta_{W}=0.226074. 
$$
From Fig.~\ref{fig:2} it is clear that corrections of a few per cent
could arise for angular distributions. They should, however, vanish after
integration over $M_{1}$ and $M_{2}$, as was mentioned before. 
The non-factorizable corrections $\delta_{nf}(M_1,M_2)$ to the double 
invariant-mass distribution are presented in Table~\ref{tab:1}. From those 
results one can expect that $\delta_{nf}$ will be less steep for 
the $M_1$ distribution than for the $M_{av}$ distribution. %
This is confirmed by Fig.~\ref{fig:3}. The corrections shown in 
Fig.~\ref{fig:3} lead to a shift in the position of the maximum of the 
Breit--Wigner curves of the order of 1--2~MeV.
These results have been obtained for the centre-of-mass energy 
$\sqrt{s}=184~\gev$. On the interval 170--190~GeV the largest corrections are
observed for 170~GeV, where the corrections are about a factor of two larger 
than those at 184~GeV. At 190~GeV the corrections are slightly smaller than 
those at 184~GeV.

\section{Conclusions}

In this letter some analytical and numerical results are presented for 
non-factorizable corrections to $W$-pair-mediated four-fermion production.
In principle these corrections could be relevant for tests of triple 
gauge-boson couplings and for the determination of the $W$-boson mass. 
For the latter the corrections are of $\cal{O}(\alpha)$ and change the $W$ 
line-shape by about 1\%. For the former they vanish at the $\cal{O}(\alpha)$
level. In view of the present experimental accuracy, the common practice
of neglecting non-factorizable corrections is justified.

One may wonder how non-factorizable corrections affect $Z$-pair-mediated and 
$ZH$-mediated four-fermion final states. In those cases, only five-point 
functions contribute, of which there are four contributions, as in 
Eq.\,(\ref{eq:HH}). However, in contrast to Eq.\,(\ref{eq:HH}), the charge
factors are pair-wise opposite, such that integration over
the decay angles leads to a vanishing result. Thus $\cal{O}(\alpha)$
non-factorizable corrections to invariant-mass distributions
in $Z$-pair-mediated or $ZH$-mediated four-fermion processes vanish.
\vspace*{0.5cm}\\

\noindent {\large \bf Acknowledgements:} 
The authors are grateful to Dr.~G.J.~van~Oldenborgh for useful discussions and 
for making some of his programs available to us.

\end{document}